\documentclass[hyper,letterpaper,12pt]{article}
\usepackage{a4wide}
\usepackage{amsmath}
\usepackage[
      colorlinks=true,
      linkcolor=blue,
      urlcolor=blue,
      filecolor=black,
      citecolor=blue,
            ]{hyperref}
\usepackage{color}
\usepackage{graphicx}
\usepackage{amsfonts}
\usepackage{amssymb}
\usepackage{epstopdf}

\def\be{\begin{equation}}
\def\ee{\end{equation}}
\def\beq{\begin{eqnarray}}
\def\eeq{\end{eqnarray}}

\begin{document}
\title{P-V criticality in the extended phase space of Gauss-Bonnet black holes in AdS space}
\author{Rong-Gen Cai$ ^1$~\footnote{email: cairg@itp.ac.cn}, Li-Ming Cao$^{2,1}$\footnote{email: caolm@ustc.edu.cn}, Li Li$^1$~\footnote{email: liliphy@itp.ac.cn}, and Run-Qiu Yang$^1$~\footnote{email: aqiu@itp.ac.cn}
\\
\\
\small $^1$
State Key Laboratory of Theoretical Physics, \\
\small Institute of Theoretical Physics, Chinese Academy of Sciences,\\
\small P.O. Box 2735, Beijing 100190, China \\
\small $^2$ Interdisciplinary Center for Theoretical Study\\
\small University of Science and Technology of China, Hefei, Anhui 230026, China }

 \maketitle

\begin{abstract}
We study the $P-V$ criticality and phase transition in the extended phase space of charged Gauss-Bonnet black holes in anti-de Sitter space, where the cosmological constant
appears as a dynamical pressure of the system and its conjugate quantity is the thermodynamic volume of the black holes. The black holes can have a Ricci flat ($k=0$), spherical ($k=1$), or hyperbolic ($k=-1$) horizon. We find that for the Ricci flat and hyperbolic Gauss-Bonnet black holes, no $P-V$ criticality and phase transition appear, while for the black holes with a spherical horizon,  even when the charge of the black hole is absent, the $P-V$ criticality and the small black hole/large black hole phase transition will appear, but it happens only in $d=5$ dimensions; when the charge does not vanish, the $P-V$ criticality and the small black hole/large phase transition always appear in $d=5$ dimensions; in the case of $d\ge 6$, to have  the $P-V$ criticality and the small black hole/large black hole phase transition, there exists an upper bound for the parameter $b=\widetilde{\alpha}|Q|^{-2/(d-3)}$, where $\tilde {\alpha}$ is the Gauss-Bonnet coefficient and $Q$ is the charge of the black hole.  We calculate the critical exponents
at the critical point and find that for all cases, they are the same as those in the van der Waals liquid-gas system.

\end{abstract}
\newpage
\tableofcontents

\section{Introduction}
Over the past years, a lot of attention has been attracted  to study thermodynamic properties of black holes in anti-de Sitter (AdS) space mainly due to the AdS/CFT correspondence~\cite{Mald,Gub,Witten}, which says that thermodynamics of black holes in AdS space can be identified with that of dual strongly coupled conformal field theory (CFT) in the boundary of the AdS space. The thermodynamic properties  of black holes in AdS space are quite different from those of black holes in asymptotically flat spacetime or in de Sitter space. In AdS space, large black holes are thermodynamically stable, while small ones are thermodynamically unstable. Under a certain temperature,
there does not exist any black hole solution in AdS space, and a so-called Hawking-Page phase transition can happen between stable large black holes and thermal gas in AdS space~\cite{Hawking1983}. This phase transition can be explained as the confinement/deconfinement phase transition of gauge field in the AdS/CFT correspondence~\cite{Witten1}

Thermodynamics of charged black holes in AdS space has also been intensively studied in the literature. In particular, in a canonical ensemble with a fixed charge of black holes, an interesting phase transition between large black hole and small black hole has been found in \cite{CEJM}. A critical point exists in the phase diagram with charge $Q$ and  temperature $T$ of the black holes. The phase transition and critical behavior of the charged black holes are quite similar to those in the van der Waals liquid-gas system.
Note that although the $Q$-$\Phi$ diagram of the charged black holes is similar to the $P$-$V$ diagram of the van der Waals system~\cite{Shen,D. Kubiznak}, where $\Phi$ is the chemical potential
conjugate to the charge $Q$ of the black hole, while $P$ and $V$ are the pressure and volume of the van der Waals system, the analogy is problematic since the charge $Q$ is
an extensive quantity and $\Phi$ is an intensive one in black hole thermodynamics, while $P$ is an intensive quantity and $V$ is an extensive one in the van der Waals system.

Recently the analogy of charged black hole in AdS space as a van der Waals system has been further enhanced by studying critical behavior of the charged AdS black hole in the
extended phase space~\cite{D. Kubiznak}, where the cosmological constant appears as the thermodynamic pressure and its conjugate quantity as a thermodynamic volume of the balck hole. It has been  showed that both systems share the same critical exponents and have extremely similar phase diagrams. This analogy has been generalized to the higher dimensional charged black holes, rotating black holes and Born-Infeld black holes in AdS space~\cite{S. Gunasekaran}. See also relevant discussions in various cases in the references~\cite{list}. For a most recent work with further interesting observations on the higher dimensional rotating AdS black hole with a single rotation parameter, see \cite{New}.

In the usual discussions of thermodynamic properties of black holes in (A)dS spaces, the cosmological constant is treated as a fixed parameter. Indeed different cosmological constant implies a different gravity theory under consideration. In the first law of black hole thermodynamics, the mass $M$, angular momentum $J$ and charge $Q$ are conserved charges, while the cosmological constant is a parameter for a given theory. However, there are some  physical reasons to view the cosmological constant as
 a variable~\cite{D. Kubiznak}.  For example,  one may suppose that there exist some more fundamental theories where some physical constants such as Yukawa coupling, gauge coupling constants, Newton's constant, and/or cosmological constant, may not be fixed values but dynamical ones arising from the vacuum expectation values.  In that case, it is natural to add these variations of ``constants'' into the first law of black hole thermodynamics~\cite{G. W. Gibbons,J. Creighton}. For example, in the gauged supergravity, the cosmological constant appears as a coupling constant.
 In addition, if the cosmological constant term is absent in the first law of black hole thermodynamics, the scaling argument cannot lead to a consistent Smarr relation for balck hole thermodynamics. Similar situation appears
 for the Born-Infeld black holes. To get a consistent Smarr relation by scaling argument, one has to introduce the Born-Infeld parameter term into the first law of Born-Infeld black holes~\cite{D. Rasheed, N. Breton}.  Once one regards the cosmological constant as  thermodynamic pressure in the first law, the black hole mass $M$ should be explained as enthalpy rather than internal energy of the system~\cite{D. Kastor}. In the geometric units $G_N =\hbar=c=k=1$,  one can identify the cosmological constant with the pressure as
\begin{equation}\label{P1}
    P=-\frac1 {8\pi}\Lambda=\frac{(d-1)(d-2)}{16\pi l^2},
\end{equation}
in $d$ dimensional spacetime. Then the thermodynamic quantity conjugate to the pressure is called  ``thermodynamic volume" of black holes~\cite{B. Dolan2,CGKP,DKKMP,Castro:2013pqa,El-Menoufi:2013pza}.

 With the new identification, one can study more thermodynamic quantities of black holes such as adiabatic compressibility, specific heat at constant pressure, or even the ``speed of sound'' associated with the black holes~\cite{B. Dolan1,B. Dolan2,B.Dolan3}. Furthermore, as pointed out by Dolan~\cite{B. Dolan2}, this also leads to an interesting possibility to reconsider the critical behavior of AdS black holes in an extended phase space, including pressure and volume as thermodynamic variables. In this way, one can more naturally consider the analogy between the AdS black holes and the van der Waals liquid-gas system. Indeed, the authors of Ref.~\cite{D. Kubiznak} initiated the investigation of the $P-V$ critical behavior of a charged AdS black hole in the extended phase space and found that the $P-V$ diagram of the black hole is exactly the same as
 the one for the van der Waals liquid-gas system. The phase transition of the charged AdS black hole has the same critical exponents as the van der Waals system. As a result
 the analogy between the charged AdS black hole and the van der Waals system becomes complete.

In this paper we are going to study the $P-V$ criticality in the extended phase space of charged Gauss-Bonnet black holes in AdS spaces. The motivations are as follows. 1) Is there any effect of some higher derivative terms of curvature on the $P-V$ criticality in the picture of \cite{D. Kubiznak,S. Gunasekaran}? 2) Note that in the discussions of
\cite{D. Kubiznak,S. Gunasekaran,list}, the electric charge $Q$ of the black holes plays an important role in the analogy, we want to know whether the charge is essential in this picture. 3) In the above discussions, the topology of the black hole horizon is a sphere. Note that in AdS space the black hole horizon could also be Ricci flat or hyperbolic~\cite{topology}. Therefore it is interesting to see whether the $P-V$ criticality appears or not for those topological black holes. Indeed we find some new features when the Gauss-Bonnet term is present.

 This paper is organized as follows. In the next section we give some thermodynamic quantities of the charged Gauss-Bonnet black holes in AdS space. In section 3, we will investigate the $P-V$ criticality of the Gauss-Bonnet black holes without charge, and compare with the van der Waals system. In section 4, we will discuss the general case with charge. Section 5 is devoted to the conclusions and discussions.

\section{Thermodynamics of Gauss-Bonnet  black holes in AdS space}

Consider the $d$-dimensional  Einstein-Maxwell theory with a Gauss-Bonnet term and a cosmological constant $\Lambda$:
\begin{equation}
S=\frac1{16\pi}\int d^dx \sqrt{-g}[R-2\Lambda+\alpha_{GB} (R_{\mu\nu\gamma\delta}R^{\mu\nu\gamma\delta}-4R_{\mu\nu}R^{\mu\nu}+R^2)-4\pi F_{\mu\nu}F^{\mu\nu}],
 \label{action}
\end{equation}
where $\alpha_{GB}$ is the Gauss-Bonnet coefficient with dimension $[{\rm length}]^2$ and the cosmological constant $\Lambda=-\frac{(d-1)(d-2)}{2l^2}$, $F_{\mu\nu}$ is the Maxwell field strength defined as $F_{\mu\nu}=\partial_\mu A_\nu-\partial_\nu A_\mu$ with vector potential $A_\mu$. In the low energy effective action of heterotic string theory, $\alpha_{GB}$ is  proportional to the inverse string tension with positive coefficient~\cite{D. G. Boulware}. Thus in this paper we will consider the case with a positive Gauss-Bonnet coefficient, namely,  $\alpha_{GB}\geq0$. In addition, let us mention here that the Gauss-Bonnet term, $R_{\rm GB}=R_{\mu\nu\gamma\delta}R^{\mu\nu\gamma\delta}-4R_{\mu\nu}R^{\mu\nu}+R^2$, is a topological term in $d=4$ dimensions and
has no dynamics in this case. Therefore we will consider $d \ge 5$ in  what follows.

The action admits a static black hole solution with metric
\begin{equation}
ds^2=-f(r)dt^2+f^{-1}(r)dr^2+r^2h_{ij}dx^idx^j,
 \label{ds1}
\end{equation}
 where $h_{ij}dx^idx^j$ represents the line element of a $(d-2)$-dimensional maximal symmetric Einstein space with constant curvature $(d-2)(d-3)k$ and volume $\Sigma_k$. Without loss of the generality, one may take $k = 1$, $0$ and $-1$, corresponding to the spherical, Ricci flat and hyperbolic topology of the black hole horizon, respectively. The metric function $f$ is given by~\cite{D. G. Boulware,RGCai2002,D. L. Wiltshir,M. Cvetic}
\begin{equation}
f(r)=k+\frac{r^2}{2\widetilde{\alpha}}\left (1-\sqrt{1+\frac{64\pi\widetilde{\alpha} M}{(d-2)\Sigma_k r^{d-1}}-\frac{2\widetilde{\alpha} Q^2}{(d-2)(d-3)r^{2d-4}}-\frac{64\pi\widetilde{\alpha} P}{(d-1)(d-2)}} \right ),
\label{fr}
\end{equation}
where $\widetilde{\alpha}=(d-3)(d-4)\alpha_{GB}$, $M$ is the black hole mass, $Q$ is related to the charge of the black hole and $P=-\frac{\Lambda}{8\pi}$. Note that in order to have a well-defined vacuum solution with $M=Q=0$,  the effective Gauss-Bonnet coefficient $\widetilde{\alpha}$ and pressure $P$ have to satisfy the following constraint
\begin{equation}
0\leq\frac{64\pi\widetilde{\alpha} P}{(d-1)(d-2)}\leq 1.
\label{al}
\end{equation}
The horizon radius $r_h$  of the black hole is determined by the largest real root of the equation $f(r_h)=0$ and the mass  $M$ can be expressed in terms of the horizon $r_h$
\begin{equation}
M=\frac{(d-2)\Sigma_k r_h ^{d-3}}{16\pi}\left (k+\frac{k^2\widetilde{\alpha}}{r_h^2}+\frac{16\pi P r_h^2}{(d-1)(d-2)}\right )+\frac{\Sigma_k Q^2}{8\pi (d-3)r_h^{d-3}}.
\label{M}
\end{equation}
The Hawking temperature of the black hole can be easily obtained by requiring the absence of conical singularity at the horizon in the Euclidean sector of the black hole solution, which is given by
\begin{equation}
T=\frac1{4\pi}f'(r_h)=\frac{16\pi P r_h ^4/(d-2)+(d-3)k r_h ^2+(d-5)k^2 \widetilde{\alpha}-\frac{2 Q^2}{(d-2)r_h ^{2d-8}}}{4\pi r_h  (r_h ^2+2k\widetilde{\alpha})}.
\label{T}
\end{equation}
Since we are going to discuss the thermodynamics of the black hole in the extended phase space by introducing the pressure $P=-\frac{\Lambda}{8\pi}$, the black  hole mass $M$  should be considered as the enthalpy $H\equiv M$ rather than the internal energy of the gravitational system~\cite{D. Kastor}. Other thermodynamic quantities can be obtained through thermodynamic identities. For examples, the entropy $S$, thermodynamic volume $V$ and electric potential (chemical potential) $\Phi$ are given by
\begin{equation}
S=\int_0^{r_h} T^{-1}(\frac{\partial H}{\partial r})_{Q,P} dr=\frac{\Sigma_k r^{d-2}_h}{4}\left (1+\frac{2(d-2)\widetilde{\alpha} k}{(d-4)r_h^2}\right ),
\label{S}
\end{equation}
\begin{equation}
V=(\frac{\partial H}{\partial P})_{S,Q}=\frac{\Sigma_k r_h ^{d-1}}{d-1},
\label{vol}
\end{equation}
\begin{equation}\label{Qphi}
    \Phi=(\frac{\partial H}{\partial Q})_{S,P}=\frac{\Sigma_k Q}{4\pi (d-3)r_h^{d-3}}.
\end{equation}
We can see that the thermodynamic volume is a monotonic function of the horizon radius $r_h$ in our case. Taking advantage of this, one can use the horizon radius to stand for the conjugate volume in the equation of state. In the later discussions, we will see that just the horizon radius corresponds to the specific volume in the Van der Waals equation rather than the thermodynamic volume.

It is easy to check that those thermodynamic quantities satisfy the following differential form
\begin{equation}\label{diff-Small}
dH= TdS +\Phi dQ +VdP +\mathcal{A} d\tilde {\alpha},
\end{equation}
where
\begin{equation}\label{beta}
    \mathcal{A} \equiv (\frac{\partial H}{\partial\widetilde{\alpha}})_{S,Q,P}=\frac{(d-2)k^2\Sigma_k}{16\pi} r^{d-5}_h- \frac{(d-2)k\Sigma_k T}{2(d-4)}r_h^{d-4}
\end{equation}
is the conjugate quantity to the Gauss-Bonnet coefficient $\tilde {\alpha}$.
  Note that here we have treated the Gauss-Bonnet coefficient as a variable. By the scaling argument, we can obtain the generalized Smarr relation for the
black holes
\begin{equation}\label{Smarr}
    (d-3)H=(d-2)TS-2PV+2\mathcal{A}\alpha+(d-3)Q\Phi.
\end{equation}
For black holes in Lovelock gravity, a generic relation is given in~\cite{Kastor:2010gq}. Note that for the Born-Infeld black hole case, the Born-Infeld parameter $b$ has units of the electric intensity and the corresponding conjugate quantity $\mathcal {B}$ has units of electric polarization, the authors of Ref.~\cite{S. Gunasekaran} therefore interpret the conjugate quantity $\mathcal {B}$ as Born-Infeld vacuum polarization.  In our case,  note that the Newton's constant $G_N$ has dimension $[\rm length]^{d-2}$, $\mathcal {A}$ in fact has dimension $[\rm length]^{-3}$.  In addition, the black hole has volume $V\sim \Sigma_k r^{d-1}_h$, area
$A\sim \Sigma_k r_h^{d-2}$,  the black hole horizon has scalar curvature $R_h \sim k/r_h^2$, and the Gauss-Bonnet term on the horizon $R_{\rm GB} \sim k^2/r^4_h$. The first term in (\ref{beta}) then has the form $VR_{\rm GB}$, while the second term $ T A R_h$.  Clearly when $k=0$, both terms vanish. In particular, it is quite interesting to note that the first term in (\ref{beta}) is just the second term in the black hole mass (\ref{M}), while the second term in (\ref{beta}) is just the second term of the black hole entropy (\ref{S}) multiplied by Hawking temperature $T$.  This results is consistent with the one given in \cite{Kastor:2010gq}. There the authors have shown that the conjugate quantity $\Psi^{(k)}$ to the Lovelock coefficient $b_k$ generally consists of three terms, which are related to mass, entropy and the anti-symmetric Killing-Lovelock potential of the black hole solution, respectively. Anyway, the conjugate quantity $\mathcal{A}$ is certainly related to the effective stress-energy tensor of the Gauss-Bonnet term $R_{\rm GB}$. It is required to further understand the physical meanings of the conjugate quantity $\mathcal{A}$ in the black hole thermodynamics in the extended phase space.

Furthermore, the Gibbs free energy and Helmholtz free energy can be obtained by Legendre transformations as
\begin{equation}
G=G(T,P,Q)=H-TS, \hspace{0.5cm} F=F(T,V,Q)=G-PV.
\label{Gibbs1}
\end{equation}
It is worth pointing out that the Helmholtz free energy $F$ is calculated by subtracting off the contribution of the background of the AdS vacuum solution. Therefore, only the case with a negative $F$  is regarded as the black hole solution being thermodynamically favored over the pure AdS vacuum solution. On the other hand, one can see that the black hole entropy \eqref{S} may be negative in the case with hyperbolic horizon ($k=-1$)~\cite{T. Clunan}. A negative entropy does not make any sense in statistical physics. As a result, in the following discussions, we will impose the following constraints
\begin{equation}
F  \le 0,\hspace{0.3cm}S\geq0,\hspace{0.3cm}r_h>0,\hspace{0.3cm}T\geq0,\hspace{0.3cm} 0\leq\frac{64\pi\widetilde{\alpha} P}{(d-1)(d-2)}\leq 1.
\end{equation}
In addition, let us note that in the case $k=-1$, one can see from the metric function $f$ (\ref{fr}) that there exists a minimal horizon radius $r_h^2 \ge 2\tilde {\alpha}$.
However, the non-negative definiteness of the black hole entropy (\ref{S}) gives a more strong constraint on the horizon radius in this case: $r^2_h \ge (2+ 4/(d-4))\tilde{\alpha}$.

 \section{$P-V$ criticality in the case with $Q=0$}

 From Eq.~(\ref{M}) we can see that the charge Q controls the lowest power term of $r_h$. So one might  expect that the critical behavior would be different between the case $Q=0$ and the case $Q\neq0$. We will first study the case $Q=0$ in this section. In this case, the equation (\ref{M}) is reduced to
\begin{equation}
H=\frac{(d-2)\Sigma_k r_h ^{d-3}}{16\pi}\left (k+\frac{k^2\widetilde{\alpha}}{r_h^2}+\frac{16\pi P r_h^2}{(d-1)(d-2)}\right ).
\label{M0}
\end{equation}
From the Hawking temperature (\ref{T}) we have the equation of state of the black holes
\begin{equation}
P=\frac{d-2}{4r_h}(1+\frac{2k\widetilde{\alpha}}{r_h^2})T-\frac{(d-2)(d-3)k}{16\pi r_h^2}-\frac{(d-2)(d-5)k^2\widetilde{\alpha}}{16\pi r_h^4}.
\label{PQ}
\end{equation}
Since we want to compare it with the van der Waals equation, we make a series expansion for the van der Waals equation with the inverse of specific volume $v$,
\begin{equation}\label{Van}
    P=\frac T{v-b}-\frac a{v^2}\approx\frac T v+\frac{bT}{v^2}-\frac a{v^2}+O(v^{-3}).
\end{equation}
Thus one can identify the specific volume $v$  with the horizon radius of the black holes as~\cite{S. Gunasekaran,D. Kubiznak}
\begin{equation}\label{sv}
    v=\frac{4r_h}{d-2}.
\end{equation}
 We see that the specific volume $v$ is proportional to the horizon radius $r_h$, therefore we will just use the horizon radius in the equation of state for the black holes  hereafter in this paper.  Of course, one can easily recover all the results in terms of $v$ by the relation~(\ref{sv}).

We know that the critical point is determined as the inflection point in the $P-V$ diagram, i.e.,
\begin{equation}
\left. \frac{\partial P}{\partial r_h}\right |_{r_h=r_{hc},T=T_c}=\left. \frac{\partial^2 P}{{\partial r_h}^2}\right |_{r_h=r_{hc},T=T_c}=0,
\label{critialpont}
\end{equation}
where we have used the subscript `c' to stand for the quantities at the critical point.  From (\ref{PQ}) we can obtain the critical temperature
\begin{equation}
T_c=\frac{10\widetilde{\alpha}(d-5)+3kr_{hc}^2(d-3)}{4\pi r_{hc}(r_{hc}^2+12k\widetilde{\alpha})},
\label{TcQ0}
\end{equation}
and the equation for the critical horizon radius (specific volume)
\begin{equation}
(d-3)r_{hc}^4-12kr_{hc}^2\widetilde{\alpha}+(12d-60)\widetilde{\alpha}^2 = 0.
\label{eqc}
\end{equation}
It can be seen that the physical solutions of equation~\eqref{eqc} depend on the dimension and topology of the black hole horizon. We will give a detailed discussion for each situation in the next subsection.
\subsection{The phase transition for different horizon topologies and dimensions}

The topology of the black hole horizon is characterized by the value of $k$, where $k=0$, $k=-1$, and $k=+1$ correspond to the Ricci flat,  hyperbolic and spherical horizon, respectively. For each case of $k$, we will investigate the phase structure and criticality in the extended phase space.

\subsubsection*{1) Ricci flat case with $k=0$}

For the Ricci flat case, i.e. $k=0$, the equation~\eqref{eqc} has no positive solution for $r_{hc}$, which means that there  does not exist any critical point. This can be seen more clearly from  the equation of state \eqref{PQ}. When $k=0$, it reduces to
\begin{equation}
P=\frac{d-2}{4r_h}T.
\end{equation}
The pressure as a function of $r_h$ is monotonic for a fixed temperature. Thus no phase transition can happen.

\subsubsection*{2) hyperbolic case with  $k=-1$}

In this case, the equation~\eqref{eqc} can be rewritten as
\begin{equation}
(d-3)r_{hc}^4+12r_{hc}^2\widetilde{\alpha}+(12d-60)\widetilde{\alpha}^2 = 0.
\end{equation}
Only when $\widetilde{\alpha}<0$ may this equation admit positive roots.  Thus
we conclude that there does not exist any phase transition and critical point in the hyperbolic case.
We can also see the fact from the equation of state of the black holes
\begin{equation}
P=\frac{d-2}{4r_h}(1-\frac{2\widetilde{\alpha}}{r_h^2})T+\frac{(d-2)(d-3)}{16\pi r_h^2}-\frac{(d-2)(d-5)\widetilde{\alpha}}{16\pi r_h^4}.
\end{equation}
Note that the fact $r_h^2 > 2\tilde{\alpha}$, the first term in the right hand side of the above equation is always positive.  When $d=5$, the pressure is a monotonic function of $r_h$, therefore there are no critical point and phase transition in this case. When $d\ge 6$, the last term in the right hand side of the above equation dominates  for the small radius case. In this case, there is only one stable phase. Again, there do not exist any critical point and phase transition.

\subsubsection*{3) spherical case with $k=1$}

\begin{figure}[h!]
\begin{center}
\includegraphics[width=0.45\textwidth]{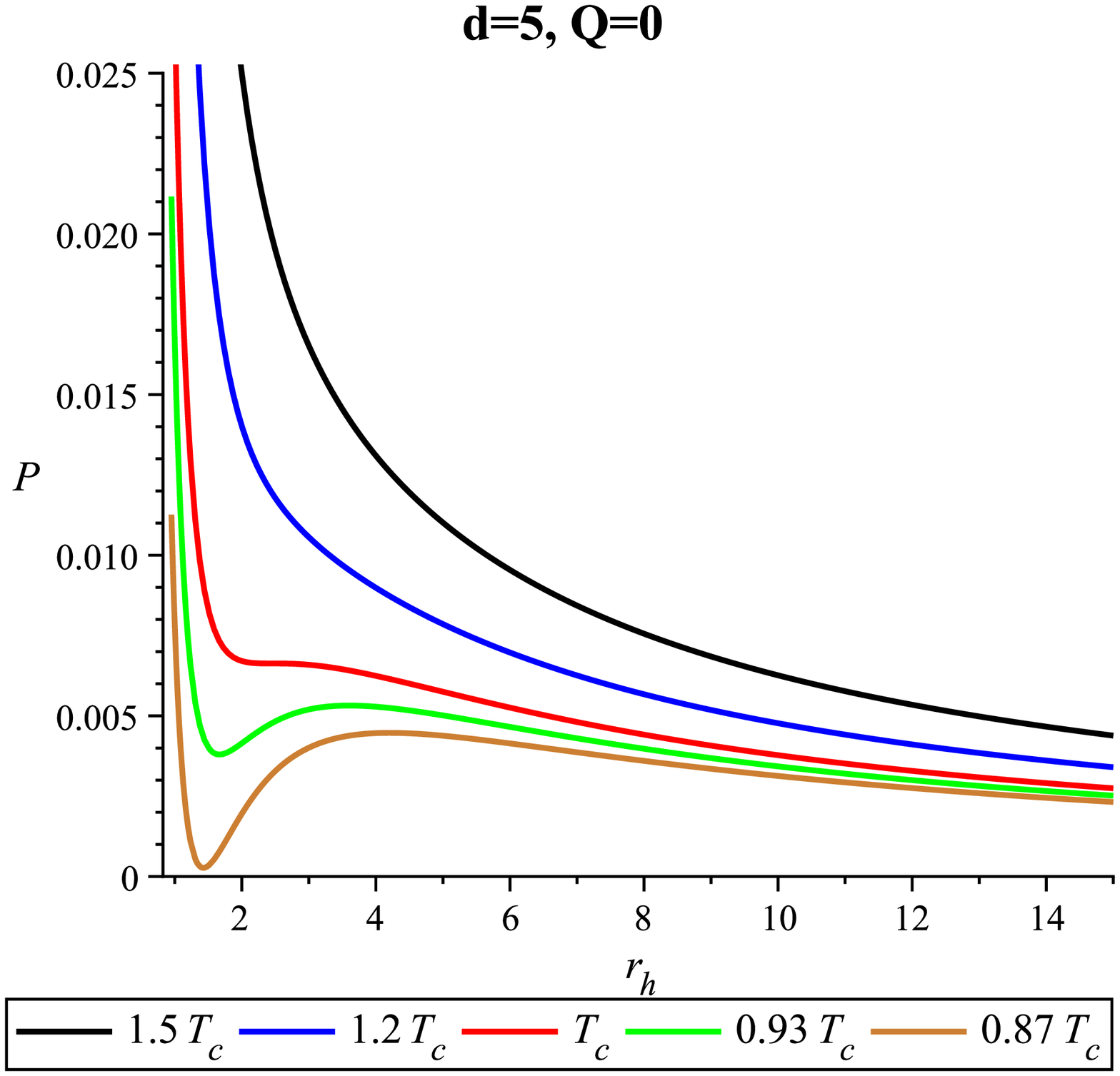}
\includegraphics[width=0.45\textwidth]{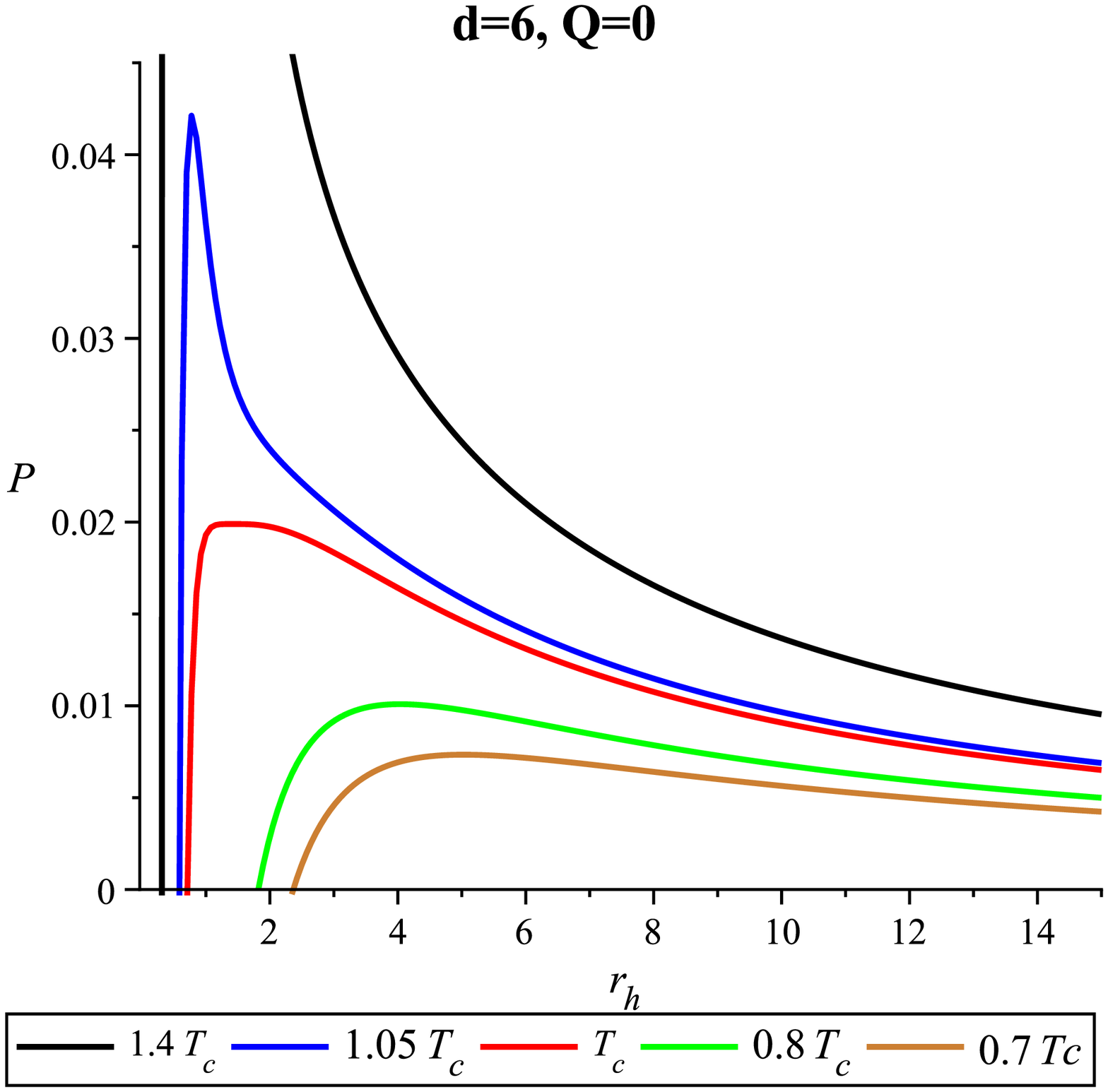}
\caption{The $P-V(r_h)$ diagram for $d=5$ (left plot) and $d=6$ (right plot) near the critical temperature, respectively. Without loss of generality, we consider the case $\widetilde{\alpha}=1$.  In both plots, the temperature of isotherms decreases from top to bottom and the pressure satisfies the relation~\eqref{al}.}
\label{fig1}
\end{center}
\end{figure}

In this case,  the equation~\eqref{eqc} reduces to
\begin{equation}
(d-3)r_{hc}^4-12r_{hc}^2\widetilde{\alpha}+(12d-60)\widetilde{\alpha}^2 = 0.
\label{eqc3}
\end{equation}
This equation has a positive real solution for $r_{hc}$ only in five dimensional and six dimensional cases. When $d=5$, we have
\begin{equation}
r_{hc}=\sqrt{6\widetilde{\alpha}},\hspace{0.5cm} P_c=\frac1{48\pi\widetilde{\alpha}},\hspace{0.5cm}T_c=\frac1{\pi \sqrt{24\widetilde{\alpha}}},
\label{d5c}
\end{equation}
from which we can easily find an interesting relation among the critical pressure $P_c$, temperature $T_c$ and horizon radius $r_{hc}$:
\begin{equation}
\frac{P_c r_{hc}}{T_c}=\frac1 4.
\label{d5c1}
\end{equation}
This relation is universal in the sense that it is independent of  $\widetilde{\alpha}$.  This result is very similar to the one in the van der Waals system, where the critical point also has an analogue universal relation with the only difference that the right hand side of equation~\eqref{d5c1} is $3/8$. From~\eqref{d5c1}, we can also see that it is more natural to view the horizon radius as the specific volume, instead of the thermodynamic volume. The corresponding $P-r_h$ diagram drawn in the left plot of figure~\eqref{fig1} is exactly the same as the $P-V$ diagram of the van der Waals liquid-gas system. The  critical isotherm is denoted by the red line in figure~\eqref{fig1}. For a fixed temperature lower than the critical one, we have two branches  whose pressure decreases as the increase of horizon radius, one is in the small radius region (corresponding to fluid phase in the van der Waals system) and the other is in the large radius region (corresponding to the gas phase). Such two branches have a positive compression coefficient, thus representing stable phases. Between them there is an unstable phase with a negative compression coefficient.  For appropriate values of pressure, the isothermal line allows two physical horizon radii. Therefore, the so-called small black hole/large black hole  phase transition occurs, which is reminiscent of the liquid/gas phase transition of the van der Waals system. Such phase transition is first order for $T<T_c$, while it becomes second order at $T_c$ just as the same as the case in the van der Waals system. Above the critical temperature, the black holes are always in the gas phase and no phase transition happens.

 In the $d=6$ case, although we can get a positive solution of the horizon radius from \eqref{eqc3}, which has a corresponding positive critical temperature $T_c$, there does not exist any phase transition below this critical temperature. As one can see from~\eqref{PQ} for $k=1$ and $d=6$, the last term now dominates the equation of state at small horizon radius, where the pressure increases with the radius. Thus, comparing to the $d=5$ case, the stable phase with small horizon radius no longer exists. This can be seen  clearly from the right plot of figure~\eqref{fig1}, which exhibits the $P-r_h$ diagram in the $d=6$ case. We see from the plot that for any fixed temperature there is only one branch with a postive compression coefficient. In other words, there do not exist two stable phases between them a phase transition could occur in this case.

\begin{figure}[h!]
\begin{center}
\includegraphics[width=0.48\textwidth]{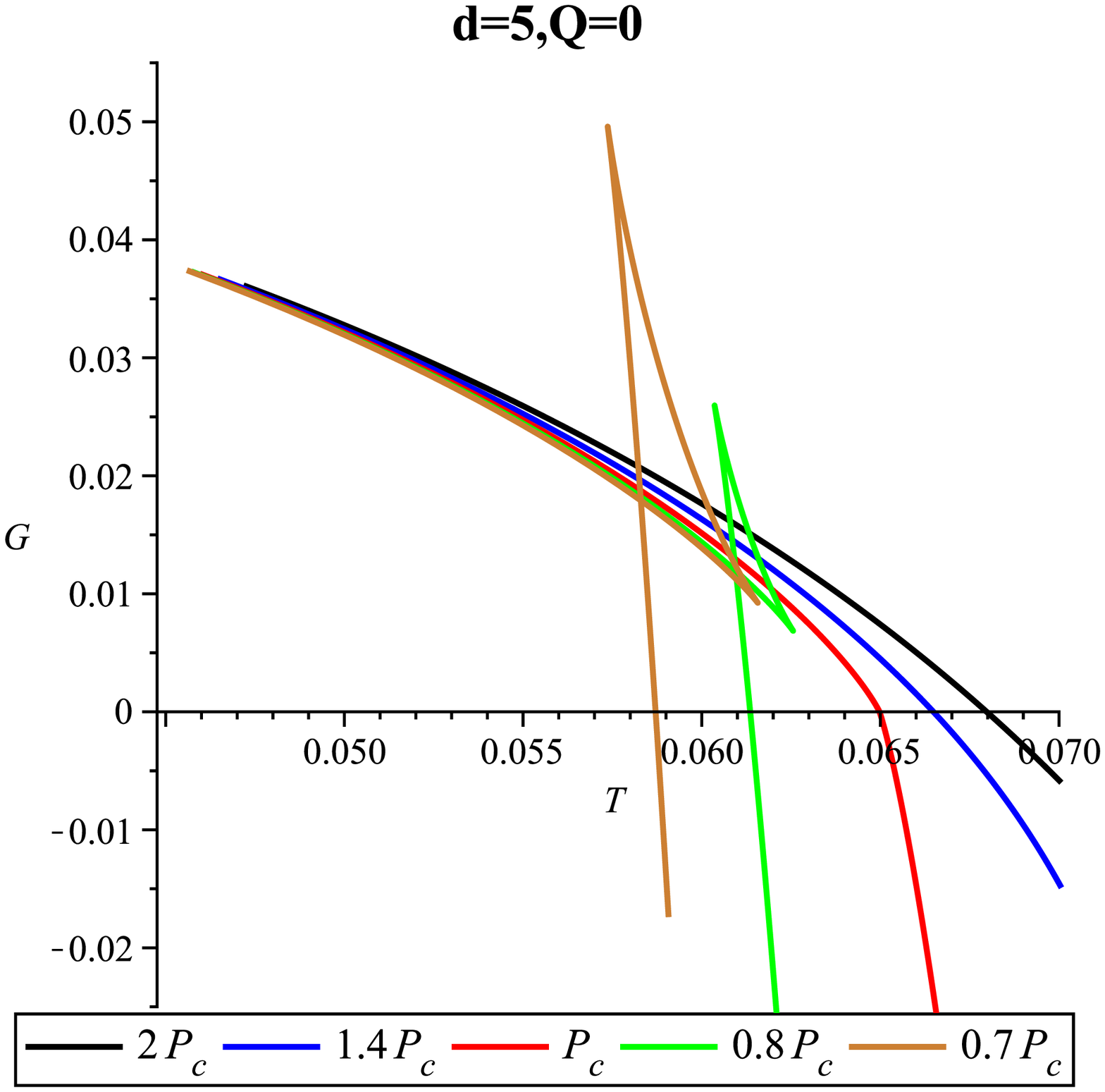}
\includegraphics[width=0.48\textwidth]{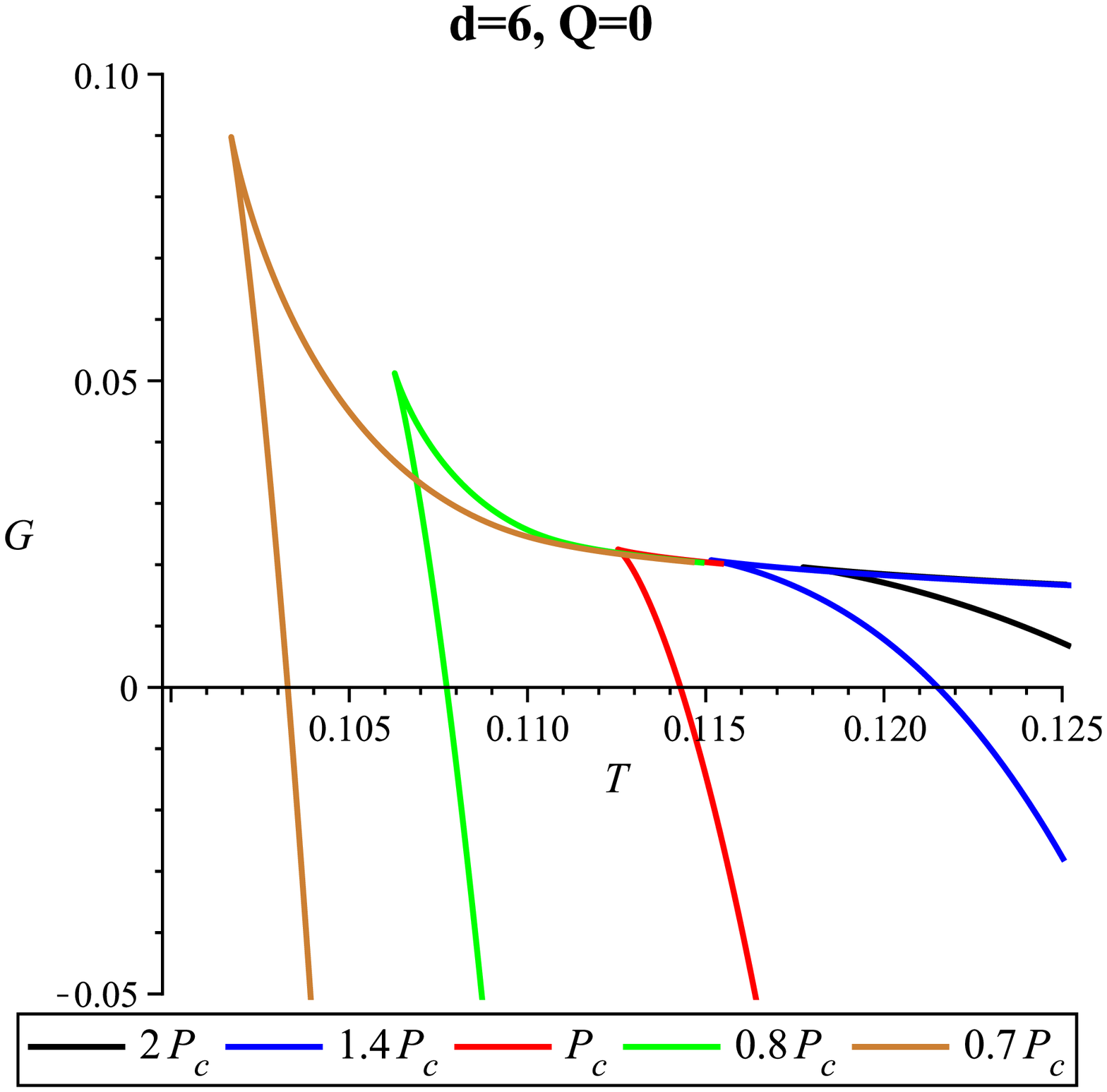}
\caption{The Gibbs free energy as a function of temperature for different pressures. The left plot is for the case with $d=5$, where the ``swallow tail" appears and
a critical point exists.  The right plot is for the case with $d=6$, where there is no phase transition. In both plots  we set $\widetilde{\alpha}=1$.}.
\label{fig2}
\end{center}
\end{figure}

We plot the Gibbs free energy defined in \eqref{Gibbs1} in figure~\eqref{fig2}  as a function of temperature for various pressures.  We can see from the left plot that in the case of $d=5$, the Gibbs free energy with respect to temperature develops a ``swallow tail" for $P<P_c$, which is a typical feature in a first order phase transition, and at $P=P_c$, the ``swallow tail"  disappears, corresponding to the critical point. On the other hand, in the case of $d=6$, there does not exist any ``swallow tail" and thus no phase transition happens.

\subsection{Critical exponents at the critical point }

From the above discussions, even when the charge is absent, we can find that the  $P-V$ criticality of a five dimensional spherical Gauss-Bonnet black hole has an impressive analogy with the $P-V$ criticality of a van der Waals liquid-gas system. This is a non-trivial result, since the previous studies without Gauss-Bonnet term in the literature indicate that there is no $P-V$ criticality for the case of vanishing charge. Here the Gauss-Bonnet coefficient $\widetilde{\alpha}$ in some sense plays the role of charge. Clearly the existence of $P-V$ criticality and the small black hole/large black hole phase transition is closely related to the existence of the small local stable Gauss-Bonnet black holes in AdS space~\cite{RGCai2002}. Note that such a stable small black hole does not exist when $d\ge 6$.

 In this subsection, we will study the critical behavior of some physical quantities and compute corresponding critical exponents.  As a warm up, we will first give a brief introduction to the critical exponents for the fluid/gas  phase transition in the van der Waals liquid-gas system before applying it to the black hole system.

Near the critical point, the critical behavior of a van der Waals liquid-gas system can be characterized by the following critical exponents~\cite{R. E. Reichl}:
\begin{eqnarray}
  C_v &\sim & (-\frac{T-T_c}{T_c})^{-\alpha}, \nonumber \\
  \frac{v_g-v_l}{v_c}&\sim & (-\frac{T-T_c}{T_c})^{\beta}, \nonumber \\
  \kappa_T&\sim & (-\frac{T-T_c}{T_c})^{-\gamma},\nonumber \\
  P-P_c&\sim & (v-v_c)^{\delta},
  \label{ce}
\end{eqnarray}
 where $T<T_c$ and $C_v=T(\frac{\partial S}{\partial T})|_v$ is the the heat capacity at constant volume and $\kappa_T=-v^{-1}(\frac{\partial v}{\partial P})|_T$ is the isothermal compressibility. The index $c$ denotes that a quantity is taken value at the critical point of the van der Waals liquid-gas system. Subscripts $g$ and $l$ stand for the gas phase and liquid phase, respectively. For the van der Waals liquid-gas system, the critical exponents take values as follows: $\alpha=0, \beta=1/2, \gamma=1, \delta=3$.

Now let us turn to calculate the critical exponents in the Gauss-Bonnet black hole system. In order to examine the critical behavior near the critical point, we introduce the following expansion parameters
\begin{equation}
\tau=\frac T {T_c}-1,\hspace{0.5cm}\epsilon=\frac{r_{h}}{r_{hc}}-1\hspace{0.5cm}p'=\frac P{P_c}.
\label{para}
\end{equation}
Then we can make Taylor series expansion for the equation of state (\ref{PQ}) with $d=5$ and $k=1$. We assume the expansion has the form as~\cite{D. Kubiznak}
\begin{equation}
p'=1+{ a_{10}}\,\tau+{ a_{11}}\,\tau\,\epsilon +{ a_{03}}\,{\epsilon }^{3}+O(\tau\epsilon^2, \epsilon^4).
\label{P2}
\end{equation}
We can obtain all the expansion coefficients in (\ref{P2}) through computing its derivatives with respect to $\tau$ and $\epsilon$ at the critical point. These coefficients turn out to be
\begin{equation}
a_{10}=4, \hspace{0.5cm} a_{11}=-6, \hspace{0.5cm}a_{03}=-6.
\label{ass}
\end{equation}
%
 Using Maxwell's equal area law we obtain the following equation
\begin{equation}
0=\int_{\epsilon_l}^{\epsilon_g}\epsilon\frac{dp'}{d\epsilon}d\epsilon\Rightarrow \frac3 4 a_{03}(\epsilon_g^4-\epsilon_l^4)+\frac1 2 a_{11}\tau(\epsilon_g^2-\epsilon_l^2)=0.
\label{eqv1}
\end{equation}
On the other hand, note that during the phase transition the pressure remains unchanged, we then  have
\begin{equation}
p'|_{\epsilon_l}=p'|_{\epsilon_g}\Rightarrow a_{11}\tau(\epsilon_g-\epsilon_l)+a_{03}(\epsilon_g^3-\epsilon_l^3)=0.
\label{eqv2}
\end{equation}
These two equations~\eqref{eqv1} and~\eqref{eqv2} will be used to calculate the critical exponents here and in the next section where the electric charge does not vanish. Now substituting these expansion coefficients in~\eqref{ass} into these two equations, we conclude that these equations have a unique non-trivial solution ($\epsilon_l\neq\epsilon_g$) only when $\tau<0$. In other words, when $T<T_c$, there are two different volumes for a same pressure. The nontrivial solution of \eqref{eqv1} and \eqref{eqv2}  with \eqref{ass} reads
\begin{equation}
\epsilon_l=-\sqrt{-\tau},\hspace{0.5cm}\epsilon_g=\sqrt{-\tau}.
\label{solut}
\end{equation}
From (\ref{ce}) we can immediately conclude that the critical exponent $\beta=1/2$.
The isothermal compressibility can be calculated as follows.
\begin{equation}
\kappa_T= \left. -\frac1 v \frac{\partial v}{\partial P} \right |_{v_c}\propto \left.-\frac1 {\frac{\partial p}{\partial \epsilon}} \right |_{\epsilon=0}=\frac1{6\tau},
\label{gamma1}
\end{equation}
which indicates that the critical exponent $\gamma=1$. In addition, it can be easily seen that $(p'-1)|_{\tau=0}=-6\epsilon^3$, which tells us $\delta=3$. And the heat capacity near the critical point has $C_v=T(\frac{\partial S}{\partial T})_{r_h}=0$,  we then have $\alpha=0$.

Thus we have obtained all the critical exponents in the $P-V$ criticality of a five-dimensional spherically symmetric Gauss-Bonnet black hole in the extended phase space by treating the cosmological constant as a pressure.  It is easy to check that these critical exponents satisfy the following thermodynamic scaling laws
\begin{eqnarray}
  & &\alpha+2\beta+\gamma=2,\hspace{0.5cm}\alpha+\beta(1+\delta)=2, \nonumber \\
& &\gamma(1+\delta)=(2-\alpha)(\delta-1),\hspace{0.5cm}\gamma=\beta(\delta-1).
\label{scaling}
\end{eqnarray}
It has been confirmed that these critical exponents associated with the $P-V$ criticality of the black hole are the same as those in the van der Waals liquid-gas system,
and they obey the same scaling laws.

\section{$P-V$ criticality in the case with $Q\ne 0$}

In this section,  we will consider the case with non-vanishing charge $Q$, i.e., the charged Gauss-Bonnet black holes in AdS space. In this case,
we have the equation of state from the temperature \eqref{T} of the black holes as
\begin{equation}
P=\frac{d-2}{4r_h}(1+\frac{2k\widetilde{\alpha}}{r_h^2})T-\frac{(d-2)(d-3)k}{16\pi r_h^2}-\frac{(d-2)(d-5)k^2\widetilde{\alpha}}{16\pi r_h^4}+\frac{Q^2}{8\pi r_{h}^{2d-4}}.
\label{PQ2}
\end{equation}
%
We see that when $d=5$, the third term identically vanishes in the pressure. Therefore the behavior of the pressure in the case of $d=5$ is different from the case with $d>5$. Thus we will discuss these two cases separately. In addition, for later discussions, it is convenient to rescale some quantities in the following way
\begin{equation}
r_h=x|Q|^{\frac1{d-3}},\hspace{0.2cm}\widetilde{\alpha}=b|Q|^{\frac2{d-3}},\hspace{0.2cm}P=p|Q|^{\frac{-2}{d-3}}, \hspace{0.2cm}T=t|Q|^{\frac{-1}{d-3}}, \hspace{0.2cm}G=g|Q|,
\label{rew}
\end{equation}
where $d$ is the dimension of the black holes and $G$ the Gibbs free energy.

\subsection{Charged black holes in five dimensions}

When $d=5$, the equation of state reduces to
\begin{equation}
P=\frac3 {4r_h}(1+\frac{2k\widetilde{\alpha}}{r_h^2})T-\frac{3k}{8\pi r_h^2}+\frac{Q^2}{8\pi r_h^6}.
\label{pq2}
\end{equation}
The critical point is determined by the equation~\eqref{critialpont}. The critical point has to satisfy the following equation
\begin{equation}
r_{hc}^6-6k r_{hc}^4\widetilde{\alpha}-5k r_{hc}^2 Q^2-18\widetilde{\alpha} Q^2=0.
\label{crai1}
\end{equation}
And the critical temperature can be written as
\begin{equation}
T_c=\frac{kr_{hc}^4-Q^2}{\pi r_{hc}^3(r_{hc}^2+6k\widetilde{\alpha})}.
\label{Tc1}
\end{equation}
Rewriting these equations in terms of dimensionless quantities in~\eqref{rew}, they become
\begin{equation}
p=\frac3 {4x}(1+\frac{2kb}{x^2})t-\frac{3k}{8\pi x^2}+\frac{1}{8\pi x^6},
\label{pq3}
\end{equation}
\begin{equation}
x_c^6-6k b x_c^4-5k x_c^2-18b=0,
\label{crai}
\end{equation}
\begin{equation}
t_c=\frac{kx_c^4-1}{\pi x_c^3(x_c^2+6kb)}.
\label{Tc2}
\end{equation}
It is easy to see  that for the Ricci flat black hole with $k=0$, the pressure is a monotonic function of the horizon radius $x$. There does not exist any phase transition and therefore no critical point exists in this case. In the following we will focus on the cases with spherical topology ($k=1$) and hyperbolic topology $k=-1$, respectively.

\subsubsection*{1) spherical case k=1}

\begin{figure}[h!]
\begin{center}
\includegraphics[width=0.48\textwidth]{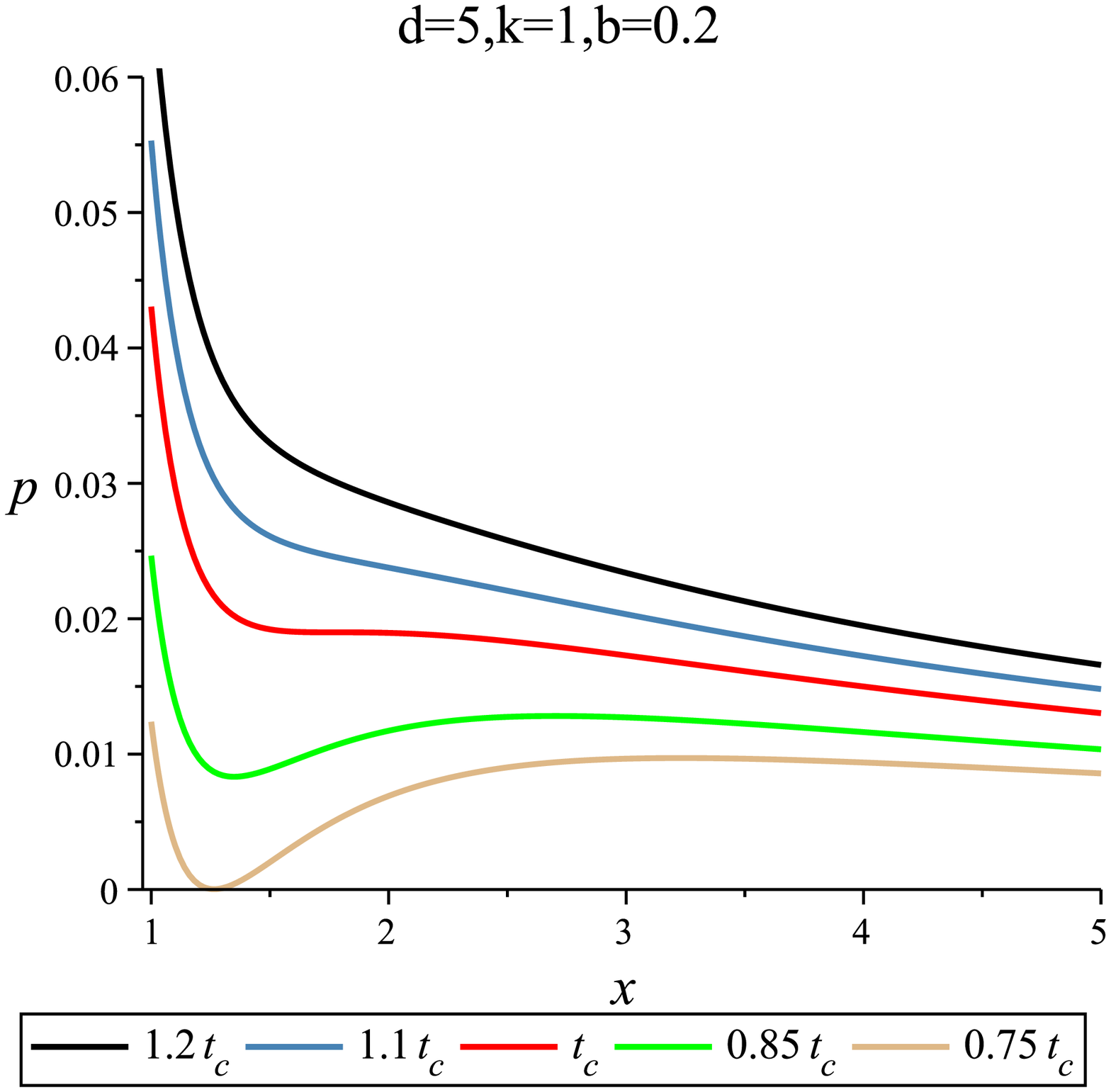}
\includegraphics[width=0.48\textwidth]{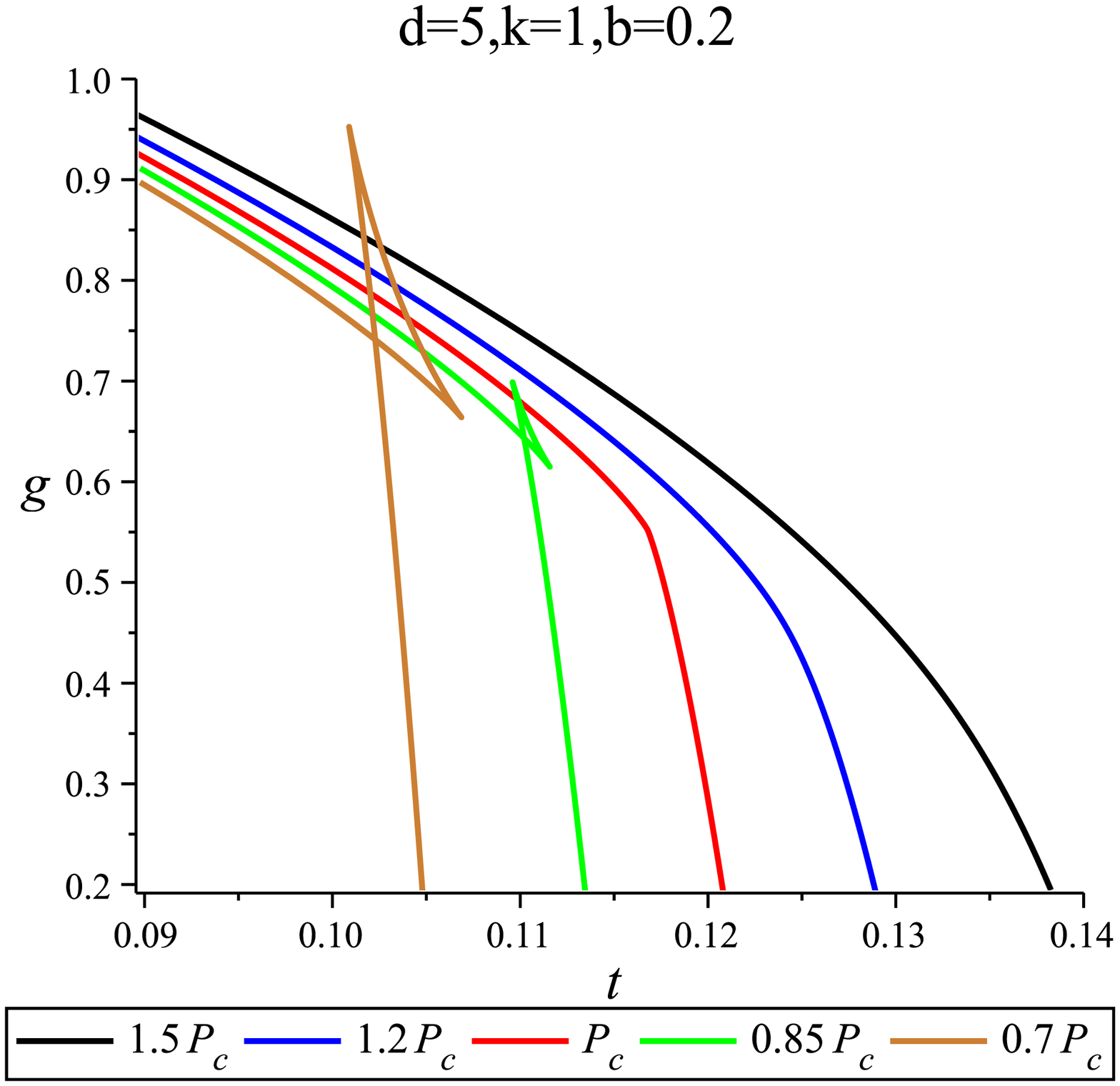}
\caption{The $p-x$ diagram (left plot) and Gibbs free energy (right plot) near the critical temperature and critical pressure for  the case of $k=1$ in five dimensions. In this figure we  take $b=0.2$. $p_c$ and $t_c$ are obtained by Eqs.~(\ref{pq3}),(\ref{crai}), and (\ref{Tc2}) with  $k=1$ and $b=0.2$.}
\label{fig3}
\end{center}
\end{figure}

In this case, the equation (\ref{crai}) reduces to
\begin{equation}
x_c^6-6b x_c^4-5 x_c^2-18b=0.
\label{crai2}
\end{equation}
Before solving this equation exactly, let us take a little time to  consider the polynomial in the left hand side of~\eqref{crai2}. We can easily get some qualitatively useful information. When $b>0$ (which means $\tilde {\alpha}>0$), the equation admits only one positive real root denoted by $x_c$. Furthermore, since the value of the polynomial at $x_c=1$(which means $r_{hc}^2=|Q|$) is $-4-24b$ which is definitely negative for $b>0$, so we can conclude that the positive real root of~\eqref{crai2} must satisfy  $x_c>1$, which leads to  a positive critical temperature (see \eqref{Tc2}).  Although this equation can have an analytic expression for the solution, we will not present it here. The $p-x$ diagram and Gibbs free energy are shown in figure~(\ref{fig3}).

In the $p-x$ diagram, the critical isotherm $T=T_c$ is denoted by the red line, where a second order phase transition happens just as the case of the van der Waals system. The curves above the critical line have a positive compression coefficient, corresponding to the ``gas" phase.  Each curve below the critical one corresponds to the case having temperature lower than $T_c$. Such curves have two stable branches, one is for small $x$ and the other is for large $x$, and the middle one between them is unstable. At these temperatures, the small black hole/large black hole phase transition occurs, which can be seen more clearly in the right plot. The Gibbs free energy as a function of temperature develops a ``swallow tail" for $P<P_c$, indicating a typical first order phase transition. The situation discussed here is comparable to the case without charge. We therefore expect that this phase transition could have the same critical exponents as the case in $Q=0$, which is indeed true as shown shortly.
\subsubsection*{2) hyperbolic case k=-1}

When $k=-1$,  the equation (\ref{crai}) reduces to
\begin{equation}
x_c^6+6 b x_c^4+5 x_c^2-18b=0.
\label{crai3}
\end{equation}
 This equation has at most one real positive root denoted by $x_c$. It can be seen that this root must be located in the range $(0, 3\sqrt{2b/5})$.
  Note that when $k=-1$, the non-negative definiteness of black hole entropy (\ref{S}) gives the constraint on the horizon radius,
\begin{equation}
r_h^2+\frac{2d-4}{d-4}k\widetilde{\alpha}\geq0.
\label{Srh}
\end{equation}
As $d=5$ and $k=-1$, the above constraint leads to
\begin{equation}
x_c^2\geq6b.
\label{Srh}
\end{equation}
With this constraint, one can see that the critical temperature (\ref{Tc2}) is always negative.  Thus we conclude that in the case $k=-1$, there does not exist any critical point and no phase transition happens. We can also obtain this conclusion from (\ref{pq3}) by noting the fact that when $k=-1$, the pressure is a monotonic function of horizon radius.

\subsection{Charged black holes in more than five dimensions}

In this case, $d\geq6$, the  equation of state of the black holes is give by (\ref{PQ2}). The critical point has to satisfy the equation
\begin{equation}
(d-3)x_c^{2d-4}-12kbx_c^{2d-6}+12(d-5)b^2x_c^{2d-8}-(4d-10)kx_c^2-12(2d-7)b=0.
\label{crcip}
\end{equation}
 And the critical temperature is given by
 \begin{equation}
 t_c=\frac{k(d-3)x_c^{2d-6}+(2d-10)bx_c^{2d-8}-2}{2 \pi x_c^{2d-7} (x_c^2+6kb)}.
 \label{Tc4}
 \end{equation}
 We can see from (\ref{PQ2}) that when $k=0$, the pressure is always a monotonic function of the horizon radius for a fixed temperature. Therefore in the case of $k=0$, the ``gas" (large black hole) phase always dominates  and it is impossible to have critical point and phase transition. We will discuss below the cases of $k=\pm 1$ separately.

 \subsubsection*{1) spherical case k=1}

 When $k=1$, one can see that the equation (\ref{crcip}) has only one positive root. The critical point, if any, has to satisfy the following constraints,
 \begin{equation}
(d-3)x_c^{2d-4}-12bx_c^{2d-6}+12(d-5)b^2x_c^{2d-8}-(4d-10)x_c^2-12(2d-7)b=0,
\label{crcip3}
\end{equation}
 \begin{equation}\label{aa3}
    (d-3)x_c^{2d-6}-2+(2d-10)b x_c^{2d-8}>0,
 \end{equation}
 \begin{equation}\label{pb2}
    0\leq64\pi b p_c \le(d-1)(d-2),
 \end{equation}
 \begin{equation}\label{Fk1}
    F|_{T=T_c,r=r_{hc}} \le 0.
 \end{equation}
The constraint (\ref{aa3}) comes from the requirement to have a positive critical temperature. And the constraint (\ref{pb2}) comes from (\ref{al}), which insures the black hole solution to be asymptotical AdS. In the case of $Q=0$, we see that the Gauss-Bonnet term plays the role of the charge in $P-r_h$ diagram.  When $b$ is very large ($b=\widetilde{\alpha}|Q|^{-2/(d-3)}\gg1$), which means $\widetilde{\alpha}$ will be much large than $Q$, one may expect that in this case, the situation is the same as the case of $Q=0$, where there does not exist a physical critical point when $d>5$. This can be seen from the pressure (\ref{PQ2}).
 So we conclude that there could exist a physical critical point only when $b$ is not too large.  This argument can also be supported by analyzing (\ref{crcip3}).
 When $b$ is very large, one has an approximate solution of the equation \eqref{crcip3} as
 \begin{equation}\label{aseq}
 x_c \sim [\frac{2d-7}{b(d-5)}]^{\frac1{2d-8}},
 \end{equation}
while the pressure (\ref{PQ2}) is dominated by the third term as
 \begin{equation}\label{doma}
 p_c\sim-\frac{(d-2)(d-5)b}{16\pi x_c^4}.
 \end{equation}
Clearly in this case the pressure is negative and the constraint \eqref{pb2} cannot be satisfied. The exact range of $b$,  in which a physical critical point exists, can be determined numerically.
 \begin{figure}[h!]
\begin{center}
\includegraphics[width=0.7\textwidth]{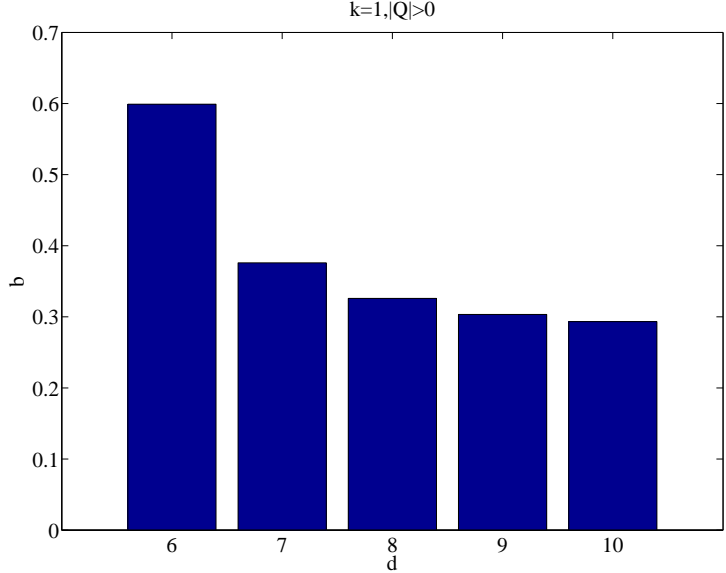}
\caption{The range of the parameter $b$ in the cases with different dimensions. In this range, a physical critical point exists. }
\label{figa}
\end{center}
\end{figure}

We plot the range of the parameter $b$ in the cases with different dimensions in figure~\eqref{figa}. We see from figure~(\ref{figa}) that when the dimension of spacetime increases, the upper bound of $b$ decreases, but finally it arrives at a finite one around $0.27$.  As a result, in the special case with $b=0$, a critical point always exists when $d \ge 6$. This is in agreement with the one for the charged AdS black holes analyzed in Ref.\cite{S. Gunasekaran}. Clearly the existence of the upper bound for the
parameter $b$ is mainly due to the competition between the third term and the fourth term of the pressure (\ref{PQ2}) in the region of small $r_h$.  When the third term dominates, the critical point is absent, while it will appear  if the fourth term dominates.  Therefore the existence of the upper bound of $b$ is the effect of the Gauss-Bonnet term.

 \subsubsection*{2) hyperbolic case k=-1}

 When $k=-1$, the physical critical point should satisfy the following constraints,
 \begin{equation}
(d-3)x_c^{2d-4}+12bx_c^{2d-6}+12(d-5)b^2x_c^{2d-8}+(4d-10)x_c^2-12(2d-7)b=0,
\label{crcip2}
\end{equation}
 \begin{equation}\label{aa2}
    [(d-3)x_c^{2d-6}+2-(2d-10)b x_c^{2d-8}](x_c^2-6b)<0,
 \end{equation}
 \begin{equation}\label{pb1}
  0\leq64\pi b p_c\leq(d-1)(d-2),
 \end{equation}
 \begin{equation}
 F|_c \le 0,
 \end{equation}
 \begin{equation}\label{aa1}
 x_c^2 \geq\frac{2d-4}{d-4}b=(2+\frac4{d-4})b.
 \end{equation}
 The former four constraints have the same origins as in the case of $k=1$. The fifth one (\ref{aa1}) comes from the requirement that the black hole entropy must be non-negative. It is easy to see that the equation~(\ref{crcip2}) has only one positive root.
Before discussing the generic case,  let us  discuss two extremal cases.

 When $b$ is very small,  we can find that the solution of (\ref{crcip2}) is $x_c\approx\sqrt{\frac{6(2d-7)}{2d-5}b}$. Both  constraints (\ref{aa2}) and (\ref{aa1}) can be satisfied. But the constraint  (\ref{pb1}) cannot be satisfied, because in this case,
 \begin{equation}\label{Pc2}
    bp_c\sim\frac{b}{8\pi x_c^{2d-4}}\sim\frac1{8\pi b^{d-3}}\rightarrow\infty,
 \end{equation}
 as $b \to 0$.  Therefore  we can conclude that when $b$ is small enough, there are no phase transition and critical point.  This is consistent with the observation that when
 $b=0$, there does not exist any real root of equation (\ref{crcip2}).

On the other hand, when $b$ is very large,  one can have an approximate solution of (\ref{crcip2}) as
$$x_c\approx \left [\frac{2d-7}{b(d-5)}\right ]^{\frac1{2d-8}}.$$
In that case, we can see that the constraint (\ref{aa1}) cannot be satisfied.

As a result, one may expect that those constraints can be satisfied only when $b$ is in some
region. However,  we can numerically check that the  real positive root of equation (\ref{crcip2}) is out of the region to have a positive Hawking temperature of the black holes.
Therefore we conclude that the critical point is also absent in this case. This is consistent with the fact that the heat capacity of the black holes at a fixed charge is always
positive in this case:
\begin{equation}
C_{P,Q}=\left ( \frac{\partial H}{\partial T}\right )_{P,Q}=\frac{(d-2)\Sigma_k r^{d-5}_h (r_h^2+2k\tilde{\alpha})T}{4} \left(\frac{\partial T}{\partial r_h}\right )^{-1}_{P,Q},
\end{equation}
where $\left(\frac{\partial T}{\partial r_h}\right )_{P,Q}$ is always positive in the physical region of the horizon radius.

\subsection{The critical behavior when $d \ge 5 $}
\begin{figure}[h!]
\begin{center}
\includegraphics[width=0.5\textwidth]{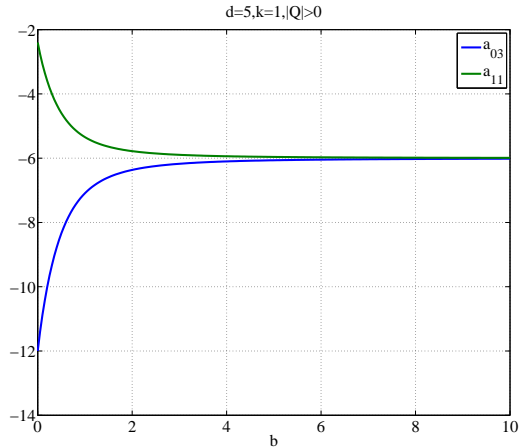}
\caption{The values of $a_{11}$ and $a_{03}$ when $d=5$, $k=1$ and $Q\neq0$. They are both negative. So the phase transition can occur when $\tau<0$.}
\label{fig6}
\end{center}
\end{figure}

We can calculate the critical exponents of some physical quantities for $Q\neq0$ in $d \ge 5$ in the same way as the case for the black holes without charge in $d=5$.  It is found that those critical exponents are universal and does not change. That is to say,  those critical exponents in the case of charged black holes in $d \ge 5$ are the exactly same as those in the case of natural black holes in $d=5$ calculated in Section 3.  We therefore will not repeat these calculations here, instead we pay some attention on some special aspects. Near a critical point, the pressure behaves as ~\eqref{P2}.  From the requirement that the pressure should be equal at transition point, we have
\begin{equation}
a_{11}\tau(\epsilon_g-\epsilon_l)+a_{03}(\epsilon_g^3-\epsilon_l^3)=0.
\label{eqv4}
\end{equation}
On the other hand, from the Maxwell's equal area law, one has
\begin{equation}
a_{11}\tau(\epsilon_g^2-\epsilon_l^2)+\frac3 2 a_{03}(\epsilon_g^4-\epsilon_l^4)=0.
\label{eqv3}
\end{equation}
Then nontrivial solution of the above two equations  appears only when $a_{11}a_{03}\tau<0$, which is given by
\begin{equation}\label{vgvl2}
    \epsilon_g=\frac{\sqrt{-a_{11}a_{03}\tau}}{3|a_{03}|} \hspace{0.5cm}\epsilon_l=-\frac{\sqrt{-a_{11}a_{03}\tau}}{3|a_{03}|}.
\end{equation}
In section 3, we have obtained the exact values of $a_{11}$ and $a_{03}$ in the case $Q=0$ and $d=5$. In  a general case, it seems impossible to give an analytical expression for  $a_{11}$ and $a_{03}$. As a result, we will numerically give two examples to show their behaviors.
Figure~(\ref{fig6}) shows the case of $d=5$, $k=1$ and $Q\neq0$. We can see from the figure that both $a_{11}$ and $a_{03}$ are negative. Thus we can conclude that the phase transition can occur when the temperature is less than $T_c$.  In addition, let us notice that when $b$ increases, both $a_{11}$ and $a_{03}$ quickly approach to $-6$, the value
in the case of $Q=0$.  This is consistent with the fact that when $b$ increases, the effect of $Q$ decays quickly by noting the relation  $|Q|=(\widetilde{\alpha}/b)$ in $d=5$ dimensions.

\begin{figure}[h!]
\begin{center}
\includegraphics[width=0.5\textwidth]{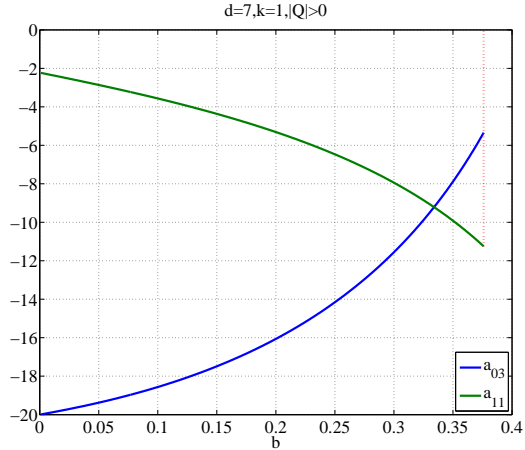}
\caption{The values of $a_{11}$ and $a_{03}$  in the case of $d=7$, $k=1$ and $Q\neq0$. In the region where $a_{11}$ and $a_{03}$  are both negative, the phase transition can occur when $\tau<0$. }
\label{fig7}
\end{center}
\end{figure}

Figure~(\ref{fig7}) shows the case of $d=7$, $k=1$ and $Q \ne 0$. We can see that in this case, the behaviors of $a_{11}$ and $a_{03}$  are different from the case of $d=5$.  Both $a_{11}$ and $a_{03}$  do not approach to a fixed value.   When $b$ is less than its upper bound, the critical point is a physical one, in this case the phase transition can occur when $T<T_c$ since $a_{11}<0$ and $a_{03}<0$;  when $b$ is beyond the upper bound, there do not exist any critical point and phase transition.

\section{Conclusions and discussions}

In this paper  we  studied the phase transition and critical behavior of topological charged Gauss-Bonnet black holes in $d$-dimensional AdS space. We discussed this issue
in the extended phase space where the cosmological constant is treated as the pressure of the thermodynamic system and its conjugate quantity is the thermodynamic volume of the
black holes. The black hole horizon can be Ricci flat ($k=0)$, spherical ($k=1$), or hyperbolic $(k=-1)$, respectively.  By scaling argument, we obtained a generalized 
Smarr relation (\ref{Smarr}) for the Gauss-Bonnet black holes. This relation implies that one has to introduce the conjugate quantity $\mathcal{A}$  to the Gauss-Bonnet coefficient $\tilde \alpha$ in the first law of black hole thermodynamics (\ref{diff-Small}). We calculated the conjugate quantity $\mathcal{A}$ and found that it consists 
of two terms, one term comes from the contribution of the Gauss-Bonnet term to the mass of the black holes, while the other from the contribution of the Gauss-Bonnet term to 
the entropy of the black holes. That is to say, one term is related to the volume of the black holes and the other is related to the area of the black holes.

We showed that for the Gauss-Bonnet black holes with a Ricci flat or hyperbolic horizon, no phase transition can appear and the large black hole phase (gas-like phase) is always dominated. This conclusion is related to the fact that for the Gauss-Bonnet black holes with Ricci flat or hyperbolic horizon, their heat capacity is always positive and 
therefore those black holes are always local thermodynamically stable~\cite{RGCai2002}. In particular, for the Ricci flat case, those thermodynamic quantities of the charged 
Gauss-Bonnet black holes are completely the same as those of the black holes without the Gauss-Bonnet term [see \eqref{M}\eqref{T}\eqref{S}]. As a result, there also do not 
exist $P-V$ criticality and small black hole/large black hole phase transition in the charged AdS black holes with Ricci flat or hyperbolic horizon in general relativity. Furthermore we believe that this conclusion is also valid for AdS black holes with Ricci flat or hyperbolic horizon in other gravity theories.

For the Gauss-Bonnet black holes with a spherical horizon, we found that even when the electric charge is absent, the $P-V$ criticality and the small black hole/large black hole phase transition can appear. It happens only in the case of $d=5$, but not for the case of $d \ge 6$.  This is related to the fact that in the $d=5$ case, there exists a local stable small black hole, while it is absent in the case of $d \ge 6$~\cite{RGCai2002}.  Also we can see from the equation of state \eqref{PQ} that when $ d \ge 6$, the last term always dominates in the region of small horizon radius, which implies that there is no branch with stable small black holes and that no phase transition can occur.
When the electric charge is not vanishing, the $P-V$ criticality and the small black hole/large black hole phase transition always appear in the case of $d=5$. Namely in this case, there is no any limitation on the parameter $b=\tilde {\alpha} |Q|^{2/(d-3)}$. While $d \ge 6$, there is an upper bound of the parameter $b$, beyond which there do not exist any  $P-V$ criticality and phase transition. The upper bound is dimensionally dependent and we obtained numerically the upper bound in some dimension cases. 
This difference between $d=5$ and $d\ge 6$ can be clearly understood from the equation of state (\ref{PQ2}). In the case of $d=5$, the third term identically vanishes and thus 
last term related to the charge of the black holes always dominates in the region of small horizon radius. Note that the last term is positive and this implies that there exists a phase with  stable small black holes. Therefore in this case the small black hole/large black hole phase transition always can happen. While in the case of $d \ge 6$, the third term no long vanishes
and thus there is a competition between the third term and last term in the region of small horizon radius. If the third term wins, there is no phase transition because 
the third term is negative and there does not exist the branch with stable small black holes in this case. On the other hand, if the last term wins, the situation behaves very like the case of $d=5$. Namely in this case, there exists the branch with stable small black holes. Thus the $P-V$ criticality and the small black hole/large black hole phase transition can appear. This is just the reason why when $d \ge 6$, there is the
upper bound on the parameter $b$. Clearly the bound on the parameter $b$ is the effect of the Gauss-Bonnet term.

In all the cases when the $P-V$ criticality appears, we calculated some critical exponents and found that those critical exponents are the same as those in the van der Waals liquid-gas system.  This further provides the supports for the analogy between the black holes in AdS space and the van der Waals liquid-gas system.

Finally let us mention that when the Gauss-Bonnet term is present, the effective AdS radius is not the one $l$ in (\ref{P1}), but~\cite{RGCai2002}
\begin{equation}
\label{l_eff}
 l_{\rm eff}= \frac{l}{\sqrt{2}}\sqrt{1+\sqrt{1-\frac{4\tilde{\alpha}}{l^2}}}.
 \end{equation}
 This can be seen by considering the vacuum solution of the theory, namely the solution (\ref{fr}) with $M=Q=0$. In this case, one may wonder whether one should define an effective pressure according to the effective AdS radius (\ref{l_eff}) through the relation (\ref{P1}) in discussing the thermodynamics of the Gauss-Bonnet black holes in the 
 extended phase space. It turns out not true. This reason is that to derive the first law (\ref{diff-Small}) of black hole thermodynamics and the generalized Smarr relation (\ref{Smarr}) in the Hamiltonian formulism where the thermodynamic volume is included as a conjugate quantity of the cosmological constant, the cosmological constant $\Lambda$ comes into the process, rather than the effective cosmological constant. See, for example, \cite{DKKMP}.

\section*{Acknowledgments}

This work was supported in part by the National Natural Science Foundation of China with grants
No.10821504,  No.11035008. No.11205148 and No.11235010.


\end{document}